\input harvmac

\def\sp{$ {\rm Sym}^k(T^{4})$}
\def\dsp{$ {\rm Hilb}^k(T^{4})$}
\def\z{Z^{\prime\prime}}
\def\symk{{\rm Sym}^k(T^4)}
\def\hilbk{{\rm Hilb}^k(T^4)}
\def\hiltil{\widetilde{\hilbk}}

\def\IL{\relax{\rm I\kern-.18em L}}
\def\IH{\relax{\rm I\kern-.18em H}}
\def\IR{\relax{\rm I\kern-.18em R}}
\def\IC{\relax\hbox{$\inbar\kern-.3em{\rm C}$}}
\def\IZ{\relax\ifmmode\mathchoice
{\hbox{\cmss Z\kern-.4em Z}}{\hbox{\cmss Z\kern-.4em Z}}
{\lower.9pt\hbox{\cmsss Z\kern-.4em Z}}
{\lower1.2pt\hbox{\cmsss Z\kern-.4em Z}}\else{\cmss Z\kern-.4em
Z}\fi}

\def\CN {{\cal N}}
\def\CR {{\cal R}}

\def\CZ {{\cal Z}}
\def\CE {{\cal E}}

\def\CH {{\cal H}}

\def\CA{{\cal A}}


\def\CN {{\cal N}}

\def\CE{{\cal E }}

\def\CZ {{\cal Z }}

\font\manual=manfnt \def\dbend{\lower3.5pt\hbox{\manual\char127}}

\def\IZ{\relax\ifmmode\mathchoice
{\hbox{\cmss Z\kern-.4em Z}}{\hbox{\cmss Z\kern-.4em Z}}
{\lower.9pt\hbox{\cmsss Z\kern-.4em Z}}
{\lower1.2pt\hbox{\cmsss Z\kern-.4em Z}}\else{\cmss Z\kern-.4em
Z}\fi}
\def\half {{1\over 2}}
\def\sdtimes{\mathbin{\hbox{\hskip2pt\vrule height 4.1pt depth -.3pt
width
.25pt
\hskip-2pt$\times$}}}

\def\CN {{\cal N}}

\def\CE{{\cal E }}

\def\CZ {{\cal Z }}


\def\IZ{\relax\ifmmode\mathchoice
{\hbox{\cmss Z\kern-.4em Z}}{\hbox{\cmss Z\kern-.4em Z}}
{\lower.9pt\hbox{\cmsss Z\kern-.4em Z}}
{\lower1.2pt\hbox{\cmsss Z\kern-.4em Z}}\else{\cmss Z\kern-.4em
Z}\fi}
\def\IB{\relax{\rm I\kern-.18em B}}
\def\IC{{\relax\hbox{$\inbar\kern-.3em{\rm C}$}}}
\def\ID{\relax{\rm I\kern-.18em D}}
\def\IE{\relax{\rm I\kern-.18em E}}
\def\IF{\relax{\rm I\kern-.18em F}}
\def\IG{\relax\hbox{$\inbar\kern-.3em{\rm G}$}}
\def\IGa{\relax\hbox{${\rm I}\kern-.18em\Gamma$}}
\def\IH{\relax{\rm I\kern-.18em H}}
\def\II{\relax{\rm I\kern-.18em I}}
\def\IK{\relax{\rm I\kern-.18em K}}
\def\IP{\relax{\rm I\kern-.18em P}}

\def\IQ{\relax\hbox{$\inbar\kern-.3em{\rm Q}$}}
\def\IP{\relax{\rm I\kern-.18em P}}

\def\inbar{\,\vrule height1.5ex width.4pt depth0pt}

\font\cmss=cmss10 \font\cmsss=cmss10 at 7pt
\def\IR{\relax{\rm I\kern-.18em R}}

\def\sdtimes{\mathbin{\hbox{\hskip2pt\vrule
height 4.1pt depth -.3pt width .25pt\hskip-2pt$\times$}}}

\def\Tr{{{\rm Tr}}}


\def\boxit#1{\vbox{\hrule\hbox{\vrule\kern8pt
\vbox{\hbox{\kern8pt}\hbox{\vbox{#1}}\hbox{\kern8pt}}
\kern8pt\vrule}\hrule}}
\def\mathboxit#1{\vbox{\hrule\hbox{\vrule\kern8pt\vbox{\kern8pt
\hbox{$\displaystyle #1$}\kern8pt}\kern8pt\vrule}\hrule}}


\def\inbar{\,\vrule height1.5ex width.4pt depth0pt}

\font\cmss=cmss10 \font\cmsss=cmss10 at 7pt
\def\IR{\relax{\rm I\kern-.18em R}}

\def\sdtimes{\mathbin{\hbox{\hskip2pt\vrule
height 4.1pt depth -.3pt width .25pt\hskip-2pt$\times$}}}


\def\makeblankbox#1#2{\hbox{\lower\dp0\vbox{\hidehrule{#1}{#2}%
   \kern -#1
   \hbox to \wd0{\hidevrule{#1}{#2}%
      \raise\ht0\vbox to #1{}
      \lower\dp0\vtop to #1{}
      \hfil\hidevrule{#2}{#1}}%
   \kern-#1\hidehrule{#2}{#1}}}%
}%
\def\hidehrule#1#2{\kern-#1\hrule height#1 depth#2 \kern-#2}%
\def\hidevrule#1#2{\kern-#1{\dimen0=#1\advance\dimen0 by #2\vrule
    width\dimen0}\kern-#2}%
\def\openbox{\ht0=1.2mm \dp0=1.2mm \wd0=2.4mm  \raise 2.75pt
\makeblankbox {.25pt} {.25pt}  }
\def\opensquare{\ht0=3.4mm \dp0=3.4mm \wd0=6.8mm  \raise 2.7pt \makeblankbox
{.25pt} {.25pt}  }

\def\sector#1#2{\ {\scriptstyle #1}\hskip 1mm
\mathop{\opensquare}\limits_{\lower 1mm\hbox{$\scriptstyle#2$}}\hskip 1mm}

\def\tsector#1#2{\ {\scriptstyle #1}\hskip 1mm
\mathop{\opensquare}\limits_{\lower 1mm\hbox{$\scriptstyle#2$}}^\sim\hskip 1mm}


\lref\fg{S. Ferrara and M. G\"unaydin, 
``Orbits of exceptional groups, duality and BPS states in 
string theory,'' hep-th/9708025}
\lref\dbr{J. DeBoer,``Large N Elliptic Genus 
and AdS/CFT Correspondence,'' hep-th/9812240}
\lref\jmone{J. Maldacena, ``The Large N Limit of Superconformal 
Field Theories and Supergravity'' Adv. Theor. Math. Phys. {\bf 2}
(1988) 231, hep-th/9711200.}
\lref\apostol{T. Apostol, {\it Modular
Functions and Dirichlet Series in Number Theory}, Springer Verlag
1990}
\lref\beauville{A. Beauville, ``Riemannian holonomy and 
algebraic geometry,'' math.AG/9902110}
 \lref\spinning{Breckenridge et. al. Spinning black holes}
\lref\carlipi{S. Carlip, ``The $(2+1)$-Dimensional Black Hole,''
gr-qc/9506079} \lref\carlipii{S. Carlip and C. Teitelboim}
\lref\deser{S. Deser, A. Gomberoff, M. Henneaux, and C.
Teitelboim, ``p-Brane Dyons and Electric-Magnetic Duality,''
hep-th/9712189} 
\lref\dvv{R. Dijkgraaf, E. Verlinde, and H. Verlinde, 
``BPS spectrum of the five-brane and black hole 
entropy,'' hep-th/9603126; ``BPS quantization of 
the five-brane,'' hep-th/9604055}
\lref\dijkgraaf{R. Dijkgraaf, ``Instanton strings
and hyperkahler geometry,'' hep-th/9810210}
 \lref\ez{M. Eichler and D. Zagier, {\it The theory of
Jacobi forms}, Birkh\"auser 1985}
\lref\tkawai{T. Kawai, ``K3 surfaces, Igusa cusp form and
string theory,'' hep-th/9710016}
\lref\kawai{T. Kawai, Y. Yamada, and S. -K. Yang,
``Elliptic genera and $N=2$ superconformal
field theory,'' hepth/9306096 }
\lref\maldacenai{J. Maldacena, ``Black holes and
D-branes,'' hep-th/9705078}

\lref\po{N.A. Obers and B. Pioline, ``U-duality 
and M-theory,'' hep-th/9809039}
\lref\ogrady{K.G. O'Grady, ``Desingularized moduli spaces
of sheaves on a K3, I \& II,'' alg-geom/9708009,
math.AG/9805099}

\lref\wite{E. Witten, Commun. Math. Phys. {\bf 109}(1987) 525;
A. Schellekens and N. Warner, Phys. Lett. {\bf B177}(1986);
Nucl. Phys. {\bf B287}(1987)317 }

\lref\cfiv{S. Cecotti. P. Fendley, K. Intriligator,
and C. Vafa, ``A new supersymmetric index,''
hep-th/9204102;
Nucl.Phys.B386:405-452,1992}

\lref\msu{J. Maldacena and L. Susskind,
``D-branes and Fat Black Holes,''
hep-th/960404;
Nucl.Phys. B475 (1996) 679-69}

\lref\dmvv{R. Dijkgraaf, G. Moore, E. Verlinde and H. Verlinde,
``Elliptic Genera of Symmetric Products and Second Quantized Strings,''
 Commun.Math.Phys. 185 (1997) 197-209}

\lref\nikulin{V.V. Nikulin, ``Integral symmetric bilinear
forms and some of their applications,'' Math.
Izvestija, {\bf 14}(1980)103}

\lref\sv{A. Strominger and C. Vafa,
``Microscopic Origin of the Bekenstein-Hawking Entropy,''
hep-th/9601029; Phys.Lett. B379 (1996) 99-104 }

\lref\trm{J.L. Petersen and A. Taormina, ``Characters of the N=4
superconformal Algebra with two central extensions,'' Nucl. Phys.
{\bf B331}(1990)556; ``Characters of the N=4
superconformal Algebra with two central extensions (II) Massless
representations,'' Nucl. Phys. {\bf B333}(1990)833}

\lref\vw{C. Vafa and E. Witten, ``A strong coupling test of S-duality,''
hep-th/9408074; Nucl. Phys. {\bf B431}(1994) 3  }

\lref\sevrin{A. Sevrin, W. Troost, and A. Van Proeyen,
Phys. Lett. {\bf B208}(1988) 447}

\lref\sen{A. Sen, `` U-duality and Intersecting D-branes,''
hep-th/9511026;  Phys.Rev. D53 (1996) 2874-2894.
}

\lref\gottschsoergel{L. G\"ottsche and W.  Soergel,
``Perverse sheaves and the cohomology of Hilbert
schemes of smooth algebraic surfaces,'' Math. Ann.
{\bf 296} (1993)235 }

\lref\harveymoore{J.A. Harvey and G. Moore, ''Algebras, BPS States, and
Strings,''
hep-th/9510182;  Nucl.Phys. B463 (1996) 315-368.}

\lref\adssss{H. Boonstra, B.   Peeters and  K. Skenderis,
``Brane intersections, anti-de Sitter Space Times and Dual Superconformal
Theories'',  Nucl. Phys. {\bf B533} (1998) 127,
hep-th/9803231;
S. Elitzur, O. Feinerman, A. Giveon and D. Tsabar,
``String Theory on $AdS_3 \times S^3 \times S^3 \times S^1$'',
 hep-th/9811245.}

\lref\vafainst{C. Vafa, ``
Gas of D-Branes and Hagedorn Density of BPS States'',
Nucl. Phys. {\bf B463} (1996) 415, hep-th/9511088;
C. Vafa,
``{Instantons on D-branes}'',
Nucl. Phys. {\bf B463 } (1996) 435,
hep-th/9512078.}

\lref\msad{J. Maldecena and A. Strominger, ``$AdS_3$ Black Holes
and the Stringy Exclusion Principle", hep-th/9804085}

\lref\kiritsis{ E. Kiritsis, ``Introduction to 
Nonperturbative String Theory'', hep-th/9708130.}


\Title{\vbox{\baselineskip12pt
\hbox{hep-th/9903163}
\hbox{IASSNS-HEP-99/16}
\hbox{HUTP-99/A006}
\hbox{YCTP-P5-99 }
}}
{\vbox{\centerline{Counting BPS Blackholes}
\smallskip
\centerline{ in}
\smallskip
\centerline{
 Toroidal  Type II String Theory}
 }}

\centerline{Juan Maldacena,$^{1,2}$ Gregory  Moore,$^{1,3}$
and Andrew Strominger$^2$ }
\bigskip\centerline{$^1$ School of Natural Sciences}
\centerline{Institute for Advanced Study}
\centerline{Princeton, NJ 08540}
\bigskip\centerline{$^2$ Department of Physics}
\centerline{ Harvard University}
\centerline{Cambridge, MA 02138}
\bigskip\centerline{$^3$ Department of Physics}
\centerline{Yale University}\centerline{New Haven, CT 06511}

\vskip .2in
\noindent
We derive a $U$-duality invariant 
formula for the degeneracies of BPS
multiplets in a D1-D5 system for toroidal
compactification of the type II string. The
elliptic genus for this system vanishes, but it is
found that BPS states can nevertheless be counted using a
certain topological partition function involving two
insertions of the fermion number operator. This is
possible due to four extra toroidal U(1) symmetries arising from
a Wigner contraction of a
large $\CN=4$ algebra $\CA_{\kappa,\kappa'}$ for $\kappa'
\rightarrow \infty$.
We also compare the answer with a counting formula 
derived from supergravity on $AdS_3\times S^3 \times T^4$ 
and find agreement within the 
expected range of validity.


\Date{}

\newsec{Introduction}

Supersymmetric indices have proven to be invaluable in the program
of accounting for black-hole entropy using D-branes \sv. In
particular, in those cases where the computation of BPS black
holes can be related to counting functions in a conformal field
theory, the elliptic genus has been of particular use.
Nevertheless, there are examples, notably toroidal
compactification of type II string, where the relevant elliptic
genus vanishes, thus giving little indication about the D-brane
BPS state degeneracies. Perhaps surprisingly, the degeneracies 
are therefore more subtle for compactification on 
$T^4$ than for $K3$. These degeneracies were first 
seriously investigated in \sen\vafainst\dvv. 
In this paper we study these degeneracies further
in the case of the three charge system of \sv\ 
consisting of $Q_1$ $D1$-branes, $Q_5$, $D5$-branes
and momentum $N$. Using a function 
closely related to the elliptic genus 
we derive $E_{6,6}(\IZ)$ $U$-dual 
expressions for the case of 
primitive charges, i.e., charges such 
that $gcd(Q_1, Q_5, N)=1$. The formula is 
given in equations $(6.2),(6.3)$ below 
and is easily derived from our   central result, 
 the  counting formula for 1/8 BPS states
given in equation (5.9)  below, valid for 
$gcd(Q_1, Q_5)=1$.

Our approach to the problem is 
to  define  an  ``index'' in the same spirit
as the
``new supersymmetric index'' of
\cfiv. These authors    investigated the
traces in supersymmetric quantum
mechanics defined by:
\eqn\nsi{
\CE_\ell = \Tr_{\CH} (-1)^F F^\ell e^{-\beta H},
}
where $H$ is the Hamiltonian and
$F$ is a fermion number operator.
For $\CN=2$ supersymmetric
theories one can take $F$ to be the
generator of a $U(1)$ invariance and
 the ``index'' with $\ell=1$ is
invariant under perturbations of $D$-terms
(but not $F$-terms).
Moreover, in general
$\CE_{\ell}$ has no special invariances for
$\ell \geq 2$. In this paper we consider the
case $\ell=2$ in the context of certain 
conformal field theories. 
 In the  problem of interest we have some extra
symmetry, namely the four U(1)
translation symmetries of the torus.
The full symmetry is a  Wigner contraction  of the
large $\CN=4$ supersymmetry algebra
$\CA_{\kappa,\kappa'}$
\sevrin . We show that the presence of this
large $\CN = 4$ algebra
leads to invariance of $\CE_{\ell=2}$ under
a class of perturbations discussed below.
 {}From the point of view of the five dimensional theory 
these indices are particular cases of  supertrace formulas \kiritsis , 
which are invariant under deformations of the theory.

\newsec{Setting the Stage }

We will consider black strings in 6D compactification
of IIB theory on $T^4$ and the black holes in
5D  compactifications on $T^5$ obtained by
wrapping these strings. In this section we summarize some 
standard facts about $U$-duality. See \po\ 
for background. 

The low energy theory of $IIB$ on $T^5$ is
given by the 32-supercharge supergravity
supermultiplet.
This has  $27$ gauge fields and $42$ scalars.
The scalar  moduli space is $E_{6,6}(\IR)/USp(8)$.
We will work in a regime of
moduli space where $T^5 = S^1 \times T^4$ is
metrically a product with a large radius for the $S^1$.
Moreover, we assume there are no Wilson lines
(of 6D gauge fields)
along the $S^1$. This submanifold of moduli space
is described by the moduli of 6D compactification
\eqn\sixdee{
\bigl[ O(5,5;\IR)/( O(5) \times O(5))\bigr] \times \IR^+
}
where the last factor is the radius of the large
$S^1$.
A subgroup of the  $U$-duality group preserving this
submanifold  is  $O(5,5;\IZ)$ ({\it not} to be confused
with the Narain duality group in 5D).

In 5D there are particles charged under the
27 gauge fields. Their charges form
 the $\IZ^{27}$ representation
of $E_{6,6}(\IZ)$.
Since the  $U$-duality symmetry is broken to
$O(5,5;\IZ)$ along \sixdee\  the  5D  particle charges
accordingly decompose as the representation:
\eqn\chrgs{
\IZ^{27} \rightarrow  II^{5,5} \oplus \IZ^{16} \oplus \IZ
}
of $O(5,5;\IZ)$. These representations have the
following interpretations.
The lattice $II^{5,5}$ is  the electric/magnetic charge
lattice of 6D strings. The representation $\IZ^{16}$
 corresponds
to the 6D charges of particles. Finally, the singlet
$\IZ$ is the momentum along the large circle.
We will denote a 5d charge vector in this decomposition
as $\gamma= (S;P;N)$.

 We are interested in
charged black holes arising from wrappings
of 6D strings on the large $S^1$,  and  
in their
BPS excitations. In the following sections we
will count these BPS excitations using a mapping
to instanton moduli space sigma models. We
will then verify that this counting is invariant
under a certain subgroup of the $U$-duality
group $E_{6,6}(\IZ)$. To explain this subgroup we need to
understand the physics of the three summands
in \chrgs.

The first summand is the charge lattice of
6D strings (general considerations show it
is a lattice, i.e., has a symmetric nondegenerate
bilinear form \deser). We can write
$II^{5,5} \cong  H_{\rm even}(T^4) \oplus II^{1,1}$.
Corresponding to the decomposition in terms of
D-branes and (fundamental strings, wrapped NS5 branes),
respectively. We can further decompose
$  H_{\rm even}(T^4) = (H_0 \oplus H_4) \oplus H_2
\cong II^{1,1} \oplus II^{3,3}$ corresponding to
a natural basis of $D1$ strings parallel to the large $S^1$,
wrapped $D5$ branes,  and wrapped $D3$-branes,
respectively.

The particle charges $P$ in 6D form the spinor
representation $\IZ^{16}$ of $O(5,5;\IZ)$. Writing
the decomposition under the $O(4,4;\IZ)$ Narain
subgroup this decomposes as
$\IZ^{16} = II^{4,4} \oplus H_{\rm odd}(T^4;\IZ)$,
corresponding to momentum, fundamental string winding,
and wrapping of D1, D3 branes. In this paper we 
often take  $P=0$. 

Now let us consider $U$-duality. 
Let us first  assume the
string charge $S\in II^{5,5}$ is a primitive
vector. It is then a standard result of lattice
theory (see, e.g. \nikulin, Theorem 1.1.2 or 
Theorem 1.14.4) that all primitive vectors
$S\in II^{5,5}$
of a given length are
equivalent under $O(5,5;\IZ)$. 
Since \nikulin\ uses some heavy machinery it 
is worth giving the following elementary example of this
phenomenon. We may identify the lattice $II^{2,2}$ 
with the set of  integral $2\times 2$ matrices. The 
signature $(2,2)$ quadratic form 
is simply the determinant. The 
$O(2,2;\IZ)$ automorphism  group acts by  left- and 
right-multiplication by $SL(2,\IZ)$: 
\eqn\lfrtact{
M:=\pmatrix{a & b \cr c & d\cr} 
\rightarrow A_L \pmatrix{a & b \cr c & d\cr} A_R
}
Now, using the standard fact that if $gcd(a,b)=1$ 
then there exist $p,q$ with $a p + bq =1 $, it is 
easy to show that $M$ can be bidiagonalized 
over $SL(2,\IZ) \times SL(2,\IZ)$ to Smith normal 
form: 
\eqn\smith{
M \cong \pmatrix{n_1 & 0 \cr 0 & n_1 n_2\cr}
}
Thus, if $M$ is primitive then the only invariant 
is the determinant, i.e., the norm-square. In a 
similar way, if $S\in II^{5,5}$ is
primitive we can, without loss of generality, put it in the form
$S=(Q_1,Q_5)\oplus \vec 0 \oplus(0,0)$ with $gcd(Q_1,Q_5)=1$
(These are the cases for which there is a sigma model description).
In other words, we can map any general string charge into
a D1-D5 system. We  then simply write $S=(Q_1,Q_5)$ and
henceforth consider the   charge vectors
\eqn\specialv{
\gamma= (Q_1,Q_5; P ; N).
}

Charge vectors of the form \specialv\
 are special because states with these
charges can be described using an instanton sigma model as in the
original discussion of \sv. It follows that invariance 
of physical quantities under $U$-duality transformations 
which preserve the form \specialv\ can lead to 
nontrivial predictions for the instanton sigma model. 
For simplicity we will henceforth consider only 
those charges $\gamma$ which can be mapped to 
the standard 3-charge system 
$\gamma = (Q_1, Q_5; \vec 0 ; N)$ of \sv. 
\foot{Whether an arbitrary charge $\gamma$ can be
so mapped is  a subtle arithmetic 
question, but the answer is probably that every 
charge vector is equivalent to a 3-charge system, 
at least if the cubic invariant is nonzero. 
The strategy for showing this 
 is the following (we have not carried out all the details). 
Using the description of  \fg\ this is equivalent to 
diagonalizability of $3\times 3$ Hermitian matrices 
over the integral split octonions $\widetilde{\bf O}_{\IZ}$ 
using $E_{6,6}(\IZ)$ transformations. It is straightforward 
to show that for any $\gamma$ there is in fact a 
3-charge system $\gamma'$ such that $\gamma \cong \gamma'$ 
$p$-adically for all $p$. Using some facts about the 
topology of $F_{4,4}$ and a result from number theory 
called the ``strong approximation theorem,'' the 
necessary Hasse-Minkowski local $\rightarrow $
global principle can be justified, so  
 the matrix can in fact be diagonalized over $\IZ$. 
We thank B. Gross for very helpful comments on 
this problem. }

The $U$-duality transformations preserving 
the 3-charge system 
$\gamma = (Q_1, Q_5; \vec 0 ; N)$
form a subgroup 
\eqn\subyoo{
(\IZ_2 \times \IZ_2) \sdtimes S_3 \subset E_{6,6}(\IZ). 
}
This group is generated by 3 transformations: 
\eqn\teezer{
\CR: (Q_1, Q_5; \vec 0 ; N) \rightarrow
(- Q_1, Q_5; \vec 0   ; - N)
}
\eqn\teeone{
\CT: (Q_1, Q_5; \vec 0 ; N) \rightarrow
(Q_5, Q_1; \vec 0   ; N)
}
\eqn\ttprime{
\eqalign{
\CT': &
(Q_1, Q_5; \vec 0 ; N) \rightarrow
(N, Q_5; \vec 0 ; Q_1) \cr}}
The transformation $\CR$ is simply a 
rotation by $\pi$ and is certainly an 
invariance of the sigma-model. 
Also, $\CT$ is 
an  order two element
of the Narain duality group $O(4,4;\IZ)$
corresponding to $T$-duality in all four
directions. 
This is supposed to be a symmetry of the conformal field
theory on the instanton moduli space. 
However $\CT'$ 
is {\it not} an invariance of the instanton sigma model.
 This is an
``STS'' type transformation in 5D which is 
not in $O(5,5;\IZ)$. 
Thus, the nontrivial predictions of 
$E_{6,6}(\IZ)$ $U$-duality for the 
instanton sigma model are reduced 
to checking invariance under 
\ttprime. 
This is what we will check below for 
degeneracies of   BPS states,
 when $Q_1,Q_5$ are relatively prime.

\newsec{The Instanton Sigma Model and its
Superconformal Symmetry}

Now we consider the standard $D1D5$ system
as  an effective string in the $05$ direction.
For the present purposes we will approximate the
 CFT for the low energy
excitations
of the theory on the string 
 by a supersymmetric sigma model \vafainst \sv:
\eqn\effcft{
\sigma\bigl[ \IR^4 \times T^4 \bigr]   \times
\sigma\bigl[{\rm Hilb}^k(T^4)\bigr].
}
Here $\sigma(X)$ denotes a supersymmetric sigma model 
with target space $X$, and  $k=Q_1 Q_5$.
The factor 
 $\sigma\bigl[ \IR^4 \times T^4 \bigr] $ is the free sigma model from the
diagonal
$U(1)$ factor in the $U(Q_1) \times U(Q_5)$ gauge
symmetry. The other degrees of freedom come from the
hypermultiplets of interacting $D1D5$ degrees of
freedom. Their target space is approximated by ${\rm Hilb}^k(T^4)$, the Hilbert
scheme of $k$ points on
$T^4$. This is a smooth resolution of the singular orbifold \sp,
and is endowed with a smooth hyperk\"ahler metric.

It is important to realize that the innocent-looking 
\effcft\ has several subtleties. First of all, 
 there should be 
an orbifold by certain translation symmetries. 
Because of a restriction to a charge zero sector, 
described below, this can be ignored. 
Furthermore, we will be working at a point in moduli space 
where the D1 branes cannot leave the fivebranes. At some
special points in moduli space, for example when all B-fields
are zero, the D1 branes  can leave the system and the CFT becomes
singular.

The  symmetries of the CFT can be deduced from standard Dbrane
technology. We assume the D1 string is in the 05 direction and the
D5 wraps the $T^4$ and 
is in the 056789 direction. The spinors, which initially
transform in the $16_+$ of the ten dimensional Lorentz group now
transform under 
\eqn\rotation{ Spin(1,1)_{05} \times \bigl[
SU(2)^+ \times SU(2)^- \bigr]_{1234} \times  \bigl[ SU(2)^+ \times
SU(2)^- \bigr]_{6789}. } 
Note that the last factor is not really a
full symmetry of the CFT since we are on $T^4$, but it is useful
to classify spinors. The ten-dimensional supersymmetries are in
the $16_+$ but only those invariant under $SU(2)^-_{6789}$
survive, i.e. only the ones with positive chirality in the 051234
directions. Thus the unbroken supersymmetry is in the
representation 
\eqn\unbrok{ (+\half; 2,1;2,1) \oplus (-\half;
1,2;2,1) } In 1+1 dimensions $\pm \half$ chiralities correspond to
left and right movers, so we see that we get (4,4) supersymmetry.
We also see that spacetime rotations in the directions
1234  act as R-symmetries of this
conformal field theory. Since we have 8 supersymmetries we can
denote the two possible multiplets as vectors and hypers. From
the center-of-mass (COM)  CFT $\sigma(\IR^4 \times T^4)$ we get a
vector and a hyper. The vector describes motion in $\IR^4$ and the
hyper describes  motion in $T^4$.

The left-moving
part of the vector multiplet  transforms  as:
\eqn\arfour{
\eqalign{
X & \in (0;2,2;1,1) \cr
\lambda & \in (+\half; 1,2;2,1)\cr}
}
and similarly for the right-moving part exchanging the  $SU(2)_{1234}$ factors
from \rotation .

The left-moving part of the hypermultiplet describing motion on
$T^4$ transforms as \eqn\arfouri{ \eqalign{ X & \in (0;1,1;2,2)
\cr \lambda & \in (+\half; 2,1;1,2)\cr} } The D1D5 strings give
hypermultiplets $(h,\psi)$ transforming as 
\eqn\arfourii{ \eqalign{ h & \in (0;1,1;2,1) \cr \psi & \in
(+\half; 2,1;1,1)\cr} } The full CFT \effcft\ has a global
$SU(2)^+_{1234} \times SU(2)^-_{1234}$ symmetry corresponding to
spacetime rotations. This is  the massive little group of
particles in 5D and will be used below to enumerate BPS
representations. The quantum numbers of the fields under this
symmetry follow from \arfour\arfouri\arfourii. Note that for  the
$\IR^4$ factor the bosons transform under the global symmetry.
Note also that all hypermultiplets \arfouri \arfourii\  transform in
the same way under $ SO(1,1) \times SU(2)^+_{1234} \times
SU(2)^-_{1234}$ and  in a different way from the vector
multiplets \arfour . This difference is what distinguishes a
vectormultiplet from a hypermultiplet in 1+1 dimensions.

 For the $T^4$ and
\sp\ factors the $SU(2)^+_{1234}\times SU(2)^-_{1234}$
 are zeromodes of
left and right-moving $SU(2)$ current algebras of
level $k$  which
are part of the left- and right-moving
$\CN=4$ superconformal algebra. In fact, in the
example of toroidal compactification there is a
larger superconformal algebra.
This arises because  there is a $U(1)^4$ current algebra
which commutes with the $SU(2)_k$, and  can be understood as follows.
The unsymmetrized product of $k$ copies of $T^4$ has four
currents which generate simultaneous translation along the four axes of
all $k$ copies of $T^4$. These four currents are permutation
invariant and therefore descend to four $U(1)$ currents in
the orbifold theory on \sp.  The resolved
conformal field theory 
on \dsp\ is determined by twenty parameters (= $4 h_{1,1}$) 
which determine the  complex structure,
Kahler class and $B$-fields \dijkgraaf\ .
Sixteen of these  are essentially associated to each $T^4$  and
the last four are involved in blowing up the orbifold points.
The values of these $20$ parameters
 are invariant under the $U(1)^4$ action.
Therefore, the $U(1)^4$ current algebra
survives  the resolution of \sp\ to \dsp.

Put more geometrically, the resolution 
$p: \hilbk \rightarrow \symk $ only depends on 
local data (such as the direction along which 
points approach each other at the orbifold 
loci) so the  obvious translation symmetry of 
$\symk$ lifts to an action of $U(1)^4$ on $\hilbk$. 

$U(1)^4$ can be regarded
as the $\kappa^\prime \to \infty$
limit of $SU(2)_{\kappa^\prime} \times U(1)$. Since the large $\CN=4$
current algebra is $SU(2)_{\kappa}
\times SU(2)_{\kappa^\prime}  \times U(1)$,
we conclude that \dsp\ conformal field theory has
a degenerate large $\CN=4$ algebra, $\CA_{\kappa ,\infty}$.
(In the following we will sometimes abuse language 
and refer to $\CA_{\kappa,\infty}$ as a large $\CN=4$ 
algebra.)

In the study of  5D black holes in 
$S^1 \times K3$ compactifications a key role was played by the
 elliptic genus for $N=2 $ conformal field theories
 defined by \wite
\eqn\egn{\CE :=
\Tr[(-1)^{2J_0^3-  2 \tilde J_0^3}
 q^{L_0} {\bar q}^{\bar L_0} y^{2J_0^3} ],}
where $J_0^3$ and $\tilde J_0^3$ are the half-integral left and right
$U(1)$ charges. Here and henceforth we normalize 
$L_0$ so that the Ramond ground states have 
$L_0=0$. The elliptic genus 
 $\CE$ is a useful object because it is
invariant under all smooth deformations of the theory.
The trace is taken in the RR sector of the conformal field theory.
Of course, it can also be defined for ${\cal N}=4$
 theories by embedding the
$U(1)$ charges in $SU(2)$.
But in theories having 
 large ${\cal N}=4$ symmetry it is not useful 
 because it always
vanishes.   We will now  show that the modified
partition function
\eqn\dft{\CE_2  :=  \Tr[(-1)^{2J_0^3-  2 \tilde J_0^3} (2 \tilde
J_0^3)^2
q^{L_0} {\bar q}^{\bar L_0} y^{2J_0^3} ],}
is an analogous topological invariant for theories with the
large $\CN=4$ symmetry.(Note that $\CE_1=0$, and indeed,  
$\Tr (J_0^3)^n = 0 $ in any $SU(2)$ representation,
for $n$ odd.)
This amounts to showing that the massive
representations of this degenerate large $\CN=4$ algebra
do not contribute to $\CE_2$.
Consider the subalgebra generated by the 
 Ramond-sector zero mode generators
$G_0^{\pm \pm}$, $Q_0^{\pm\pm}$, $J_0^3$ and  $L_0$.\foot{ Our notation is
as follows.
$G_0^{\pm \pm}$ are the supercharges, and the first $\pm$ superscript
indicates
the charge under $J_0^3$ of  $SU(2)_k$.
 $Q_0^{\pm\pm}$ are the fermionic partners of the
$U(1)^4$ current algebra.
 $( G^{\pm\pm}_0 )^\dagger = G^{\mp \mp}_0 $ and
$( Q^{\pm\pm}_0 )^\dagger = Q^{\mp \mp}_0 $.}
Since $L_0$ commutes with the
rest of the generators we can just think of it as a c-number.
The relevant commutation relations are
\eqn\commut{\eqalign{
 \{G_0^{++} , G_0^{--} \} = 2 L_0 &, ~~~~ \{G_0^{+-},G_0^{-+}\} = 2L_0
\cr
\{Q_0^{++} , Q_0^{--} \} = 1 &, ~~~~ \{Q_0^{+-},Q_0^{-+}\} = 1,\cr
[J_0^3, G_0^{\pm-} ] =\pm {1 \over 2}G_0^{\pm-} &, ~~~~ [J_0^3, G_0^{\pm+} ]
=\pm {1 \over 2} G_0^{\pm+} \cr
[J_0^3, Q_0^{\pm-} ] =\pm {1 \over 2}Q_0^{\pm-} &, ~~~~ [J_0^3, Q_0^{\pm+} ]
=\pm {1 \over 2} Q_0^{\pm+}. \cr}}
The rest of the commutators, including those  of $G$'s with  $Q$'s, vanish
if we consider states neutral with respect to $U(1)^4$, $i.e.$
with no momentum or winding on $T^4$. The general case will be discussed
momentarily.

For a massive representation, by definition $L_0 >0$. This implies
that the commutation relations of the $G$'s and $Q$'s are those of
fermionic creation and annihilation operators. We have four
creation operators $b_i^\dagger$ which we choose to have $J_0^3 =
1/2$.  The annihilation operators  then have $J_0^3=-1/2$. Let
$|0,j \rangle $ denote the state that is annihilated by all the
annihilation operators and obeys $J_0^3|0,j \rangle=j|0,j \rangle
$ for some $j$. Acting with the creation operators  we get four
states with $J_0^3 = j+1/2$, six states with $J_0^3= j+1$, four
with $J_0^3=j +3/2$ and one with $J_0^3 = j+2$. The fermion
numbers of these states alternate. It is easy to check that the
traces over this zero mode representation $\Tr_j(-1)^F = \Tr_j
(-1)^{2J_0^3}$ as well as  $\Tr_j(-1)^{2J_0^3} J_0^3$ vanish.
 One also finds by direct computation
\eqn\rth{\Tr_j(-1)^{2J_0^3} (J_0^3)^2 \propto  j^2-4(j+{1\over 2})^2
+6(j+1)^2-4(j+{3\over
2})^2
+(j+2)^2=0 .}
We conclude the massive representations do not contribute to $\CE_2$.

If we now relax the assumption that the $U(1)^4$ charges vanish,
then the
anti commutation relations of the $G$'s and $Q$'s (denoted collectively
as $b_i, b_i^\dagger$, $i=1,\cdots, 4$) are of the form
\eqn\commb{
 \{ b_i, b_j\} =0, ~~~ \{ b_i^\dagger, b_j^\dagger\} =0, ~~~
 \{b_i, b_j^\dagger \} = M_{ij}
} where $M_{ij}$ is an Hermitian  matrix which depends on $L_0$ and
the four $U(1)$ charges. We can diagonalize $M$ by a unitary
transformation. If the eigenvalues of $M$ are all non-zero, then
the $b$'s are usual creation and annihilation operators and the
trace of $(-1)^{2 J_3} J_3^2 $ vanishes. This is the case when
$L_0 > \sum_{i=1}^4 u_i^2 $ where $u_i$ are the eigenvalues of the
four U(1) charges (appropriately normalized). If $M$ has zero
eigenvalues, this is no longer the case. This happens when $L_0 =
\sum_{i=1}^4 u_i^2 $. It would be very interesting to understand
these BPS states carrying additional charges. In this paper,
however, we concentrate on the case where all these charges are
zero.

For non-degenerate large $\CN=4$ algebras $\CA_{\kappa,\kappa'}$
 with finite $SU(2)$ levels
$\kappa $ and $\kappa^\prime$, the commutators of $G_0$ and $Q_0$
have $SU(2)_\kappa\times SU(2)_{\kappa^\prime}$
current algebra zero modes on the right hand side. This complicates the
preceding argument. However in this case one may conclude from direct
examination of formulae in \trm\ that the massive characters
do not contribute to $\CE_2$.
This implies that the index \dft\ will be useful to analyze the
conformal field theory related to $AdS_3 \times S^3 \times S^3 \times
S^1 $ \adssss . In fact it would be very interesting to
compute the supergravity result since it could teach us
something about the dual conformal field theory.

As we shall see shortly, the massless characters with $L_0=0$
 do contribute
to $\CE_2$. This contribution is independent of the continuous parameters
describing the resolution of \sp\ to \dsp. Hence we can compute
$\CE_2$  for all cases from the limiting case of \sp.

\newsec{Counting spacetime BPS  multiplets}

In this section we explain the spacetime interpretation of \dft.
The $D1D5$ system on $S^1 \times T^4$ and its excitations describe
particles in 5 dimensions.
 These all transform in representations of the 5d
Poincar\'e  supersymmetry algebra with 32 real supercharges. 
The  different representations
can be characterized by the
 $Spin(4)_{1234}$ characters
\eqn\character{
\chi(y,\tilde y) :=
\Tr_{\CH_{{\rm little}} } (-1)^{F_{spacetime}}
y^{2 J_0^3} \tilde y^{2 \tilde J_0^3}
}
of the representation of the little superalgebra.
The long representations built with 32 active (i.e. broken)
 supercharges have
character
\eqn\chari{
\chi_{0/32}(y,\tilde y) =\chi_{j_L}(-y)   
\chi_{j_R}(- \tilde y) (y^{1/2} -
y^{-1/2})^8
(\tilde y^{1/2} - \tilde y^{-1/2})^8 .
}
Here the subscript indicates the number of preserved
supercharges,  $(j_L, j_R)$ are arbitrary half-integral spins,
and
\eqn\sutwo{
\chi_j(y) = y^{-2j} + y^{-2j +2 } + 
 \cdots + y^{2j} = {y^{2j+1} - y^{-2j-1}
 \over  y-y^{-1} }.
}

The BPS  states we will encounter in the $D1D5$ system
come in three kinds of short representations:

A. $M = Z_1$. $M \not=  \vert Z_i\vert, i>1 $, where
$Z_i$ are the skew eigenvalues of the central charge
matrix.
The characters are
\eqn\charvi{
\chi_{4/32}^+ = \chi_{j_L}(-y)   \chi_{j_R}(- \tilde y) 
 (y^{1/2} - y^{-1/2})^8
(\tilde y^{1/2} - \tilde y^{-1/2})^6
}

B. If instead  $M=-Z_1$ we get
\eqn\charvii{
\chi_{4/32}^- = \chi_{j_L}(-y)   \chi_{j_R}(- \tilde y)  (y^{1/2} - y^{-1/2})^6
(\tilde y^{1/2} - \tilde y^{-1/2})^8
}

C. Finally, a shorter  representation has 
character
\eqn\charviiii{
\chi_{8/32} = \chi_{j_L}(-y)   \chi_{j_R}(- \tilde y)  (y^{1/2} - y^{-1/2})^6
(\tilde y^{1/2} - \tilde y^{-1/2})^6.
}
U-duals of massive Dabholkar-Harvey states
 turn out to be in  representations of type C. 
There are also other BPS states in other representations for example
$1/2$ BPS states, etc.

We now discuss how these characters show up in 
CFT partition functions. In general for the 
CFT $\sigma(X)$ we denote
\eqn\dfzx{
Z(X) = Z_{\sigma}(q,y, \bar q, \tilde y)
 :=
\Tr_{RR } (-1)^{2J_0^3 - 2 \tilde J_0^3}
q^{L_0}
\bar q^{\tilde L_0} y^{2 J_0^3} \tilde y^{2\tilde J_0^3}
}
For the conformal field  theory \effcft\ 
this trace  is a product\foot{Again,
 we ignore a discrete translation 
orbifold action.} of two factors: One for the
COM
degrees of freedom and one  for the CFT
$\sigma({\rm Hilb}^k(T^4))$.  The  first
 factor  can be computed straightforwardly in
terms of oscillators using the
quantum numbers \arfour\arfouri\arfourii.
We will discuss the second factor in section five.

The trace \dfzx\ for the CFT \effcft\ can be 
decomposed in terms of the characters of the 
massive little superalgebra: 
\eqn\mainclaim{
\eqalign{
Z_{\sigma}(q,y, \bar q, \tilde y)  & =
\sum_{j_L, j_R}  \chi_{8/32}(y, \tilde y)  D_{8/32}(Q_1,Q_5; j_L, j_R) \cr
+
& \sum_{N=1}^\infty
\sum_{j_L, j_R} q^N \chi_{4/32}^+(y, \tilde y)  D_{4/32}^+(Q_1,Q_5,N; j_L, j_R)
\cr
 +
& \sum_{\bar N=1}^\infty
\sum_{j_L, j_R} \bar q^{\bar N}
 \chi_{4/32}^-(y,\tilde y)
D_{4/32}^-(Q_1,Q_5,\bar N; j_L, j_R) \cr
  +
 \sum_{\Delta, \bar \Delta > 0 } &
\sum_{j_L, j_R} q^\Delta \bar q^{\bar \Delta}
\chi_{0/32}(y, \tilde y)
D_{0/32}(Q_1,Q_5,\Delta, \bar \Delta ; j_L, j_R) \cr}
}
where $(\Delta,\bar \Delta)$ run over the massive
spectrum of the CFT \effcft.

In \mainclaim\ the $D$'s measure the degeneracies
of various types of representations of the
spacetime $D=5, \CN=4$ superalgebra. In particular,
$D_{8/32}(Q_1,Q_5; j_L, j_R)$ is the number of
BPS multiplets of charge $(Q_1, Q_5, N=0)$ in the
representation \charviiii . 
$D_{4/32}^+(Q_1,Q_5,N; j_L, j_R)$ is the number of
BPS multiplets of charge $(Q_1, Q_5, N>0)$ in the
representation \charvi. These are macroscopically
black holes with positive horizon area, etc.

Note that part of the structure of \mainclaim\ as a function
of $y, \tilde y$ follows from the representation
theory of the algebra $\CA_{\kappa,\infty}$. From the
COM sigma model we have an overall factor of
$(y^{1/2} - y^{-1/2})^4
(\tilde y^{1/2} - \tilde y^{-1/2})^4$.
Then, in the \sp\ sigma model we have
$\Tr(-1)^F F^{\ell} = 0 $ for $\ell=0,1$ and therefore
there is an extra factor of
$(y^{1/2} - y^{-1/2})^2
(\tilde y^{1/2} - \tilde y^{-1/2})^2$
coming from this piece. For massive reps
$\Delta >0$ of $\CA_{\kappa,\infty}$ we
showed in section three that
in fact $\Tr(-1)^F F^{\ell} = 0 $ for $\ell=0,1,2,3$
and therefore reps with $\Delta >0, \bar \Delta =0$
give a factor  of $(y^{1/2} - y^{-1/2})^4
(\tilde y^{1/2} - \tilde y^{-1/2})^2$, etc.

In order to give a counting formula for BPS
multiplets we should take a derivative of 
\mainclaim\ by 
${ 1 \over  6!} \bigl( {d \over  d\tilde y} \bigr)^6$
at $\tilde y=1$. From the CFT of the sigma model
$\sigma(\IR^4\times T^4) \times \sigma(\symk)$
  we need 4 derivatives
to act on the COM part of the sigma model and 2
derivatives to act on the $S^k(T^4)$ part. There
is a surprising cancellation of the COM
contributions from $\IR^4$ and $T^4$ 
 after setting $\tilde y = 1$ and the
$D1D5$ CFT  gives simply:
\eqn\instsig{
(y^{1/2} - y^{-1/2})^4
\half \bigl( {d \over  d\tilde y} \bigr)^2\vert_{\tilde y=1} 
Z(S^k(T^4)). 
}
Comparing with \mainclaim\ we finally obtain
the desired counting formula for representations:
\eqn\finalform{
\eqalign{
\half \bigl( {d \over  d\tilde y} \bigr)^2\vert_{\tilde y=1}
&
Z(S^k(T^4))  =   \qquad\qquad\cr
 (y^{1/2} - y^{-1/2})^2
\sum_{j_L, j_R} \chi_{j_L}(-y)  &
  (-1)^{2j_R} (2j_R + 1)  D_{8/32}(Q_1,Q_5; j_L, j_R) \cr
 + (y^{1/2} - y^{-1/2})^4
\sum_{N=1}^\infty
\sum_{j_L, j_R} q^N \chi_{j_L}(-y) &
  (-1)^{2j_R} (2j_R + 1)
  D_{4/32}^+(Q_1,Q_5,N; j_L, j_R) \cr}
}

\newsec{Computation of $\CE_{\ell=2}$}

In this section we evaluate $\CE_2$
 more explicitly.
In \dmvv\ a general formula was derived relating the partition function for
a conformal field theory with target $X$ to that of a
conformal field theory whose target is
the orbifold ${\rm Sym}^k(X)$.
The partition function for a single 
copy of $X$ defines the degeneracies 
$c(\Delta,{\bar \Delta}, \ell,
{\tilde \ell})$ via: 
\eqn\single{{ Z}\bigl( X \bigr)
=
\sum_{\Delta ,{\bar \Delta}, \ell, {\tilde \ell}} 
c(\Delta,{\bar \Delta}, \ell,
{\tilde \ell})
 q^{\Delta} {\bar q}^{\bar \Delta} y^{\ell} {\tilde y}^{\tilde \ell}.
}
Here the trace is in the RR sector. The spectrum of
$U(1)$ charges $\ell,\tilde \ell$ is integer or half-integer,
according to the parity of the complex dimension of $X$ and $\Delta,\bar \Delta$
runs over the spectrum of $L_0, \bar L_0$. The values of
$\Delta,\bar\Delta$ are in general arbitrary nonnegative  real
numbers, although the
 difference $\Delta - \bar \Delta$ is integral.

In terms of $c$,  the partition function over ${\rm Sym}^k(X)$  may be derived
using a small modification of the
discussion in \dmvv, and is:
\eqn\symm{
{\cal Z}(p,q, \bar q, y,\tilde y)
:=  \sum_{k=0}^{\infty}  p^k {Z} \bigl({\rm Sym}^k(X) \bigr)=
\prod_{n=1}^\infty  {\prod_{\Delta,{\bar \Delta}, \ell, {\tilde \ell}}}'
{ 1 \over \left( 1 - p^n  q^{\Delta\over n} {\bar q}^{{\bar \Delta} \over n}
 y^{\ell} {\tilde y}^{\tilde \ell}
\right)^{c(\Delta,{\bar \Delta}, \ell, {\tilde \ell})}
}
}
where the prime on the product indicates
that $\Delta,{\bar \Delta}$ are restricted so
that  $ {\Delta-{ \bar \Delta} \over n}$ is an integer.

We now specialize to a target space such that
$Z(X) \vert_{\tilde y=1} = Z(X)' \vert_{\tilde y=1} =0$
(as, for example, in the case $X=T^4$ due to fermion
zero modes.).
Thus we have
\eqn\sumrules{
\eqalign{
\sum_{\tilde \ell} c(\Delta,{\bar \Delta}, \ell, {\tilde \ell}) & =0 \cr
\sum_{\tilde \ell} {\tilde \ell}   c(\Delta,{\bar \Delta}, \ell, {\tilde \ell})
& =0 \cr}
}
Moreover,
we assume that the conformal field theory for $X$
has a realization of the superconformal algebra
$\CA_{\kappa,\kappa'}$ or its $\kappa' \rightarrow \infty$ contraction.
In this case we  may use the results of the
previous section to obtain the  identity
\eqn\sumsq{
\sum_{\tilde \ell} {\tilde \ell}^2
   c(\Delta,{\bar \Delta}, \ell, {\tilde \ell})   =0
 \qquad {\rm for} \qquad
 \bar \Delta>0
}
(Recall that we are taking the $U(1)$ charges to
be zero.)

Let us now compute ${\cal Z}''$.
It follows from \sumrules\ that $\cal Z$ in \symm\ is equal to one for
$\tilde y =1$. Differentiating with respect to ${\tilde y}$ gives
\eqn\first{
\partial_{\tilde y} {\cal Z} =
\left[ \sum_{n,\Delta,{\bar \Delta}, \ell, {\tilde \ell}}
 { {\tilde \ell} c(\Delta,{\bar \Delta}, \ell, {\tilde \ell})
 p^n  q^{\Delta\over n} {\bar q}^{{\bar \Delta} \over n}
 y^{\ell} {\tilde y}^{\tilde \ell-1 }
\over \left( 1 - p^n  q^{\Delta\over n} {\bar q}^{{\bar \Delta} \over n}
 y^{\ell} {\tilde y}^{\tilde \ell}
\right)
} \right] {\cal Z}
}
If we set ${\tilde y } = 1$ we get zero by \sumrules.
Next we compute the second derivative of \symm\ with respect to
${\tilde y}$ and set $\tilde y =1$.
If the second derivative acts on the ${\cal Z}$ factor in
\first\ the result vanishes when
we set ${\tilde y} =1 $. So the second derivative must act on the sum in \first
{}.
After setting ${\tilde y } =1$ one finds
that the sum over ${\bar \Delta}$ drops out, since  only the
 ${\bar \Delta}=0 $ term contributes by \sumsq.
This implies that $\Delta$ is integral and
divisible by $n$:  $\Delta = n m$ with $m=0,1,2 \cdots$.
So we get 
\eqn\second{
\half
\partial^2_{\tilde y} {\cal Z}|_{{\tilde y} =1 } =
\sum_{n\geq 1} \sum_{m\geq 0} \sum_{ \ell\in \IZ }
 { {\hat c}(n m,\ell) p^n  q^{m}
 y^{\ell}
\over \left( 1 -
 p^n  q^{m} y^{\ell}
\right)^2
}
}
Here  we have defined
\eqn\chat{
{\hat c}(\Delta ,\ell) := \half
\sum_{{\tilde \ell}} {\tilde \ell}^2 c(\Delta,0,\ell,{\tilde \ell} )
}

Expanding \second\ yields
\eqn\final{
\half \partial^2_{\tilde y} {\cal Z}|_{{\tilde y} =1 }
 = \sum_{s,n,m,\ell} s  \left(  p^n  q^{m}
 y^{\ell} \right)^s  {\hat c}(n m,\ell)
}
where $s,n\geq 1, m\geq 0, \ell\in \IZ$. 
Collecting powers of $p,q,y$ we finally obtain
our counting formula:\foot{
The counting of $D_{8/32}$ nicely matches the
known degeneracies of Dabholkar-Harvey states
\sen \vafainst . This can be shown in the present context
using the result of \gottschsoergel, (equation (7.6) 
below).}
\eqn\deg{
\eqalign{
  (y^{1/2} - y^{-1/2})^4
\sum_{j_L, j_R}  \chi_{j_L}(-y) &
  (-1)^{2j_R} (2j_R + 1)
  D_{4/32}^+(Q_1,Q_5,N; j_L, j_R)\cr
= \sum_{\ell} y^{\ell}
&
 \sum_{s \vert Q_1 Q_5, s\vert N, s\vert \ell}
s \, \hat  c \bigl( {Q_1 Q_5 N \over s^2}, {\ell \over s} \bigr) \cr}
}
We stress that this formula is only applicable
for $gcd(Q_1, Q_5) = 1$ (i.e., for a primitive vector
in the string charge lattice) because otherwise the possibility
of bound states at threshold obscures the relationship between the
sigma model and the $D1D5$ system.\foot{Thus this is 
 the formula for the BPS degeneracies only for a
fraction,  $6/\pi^2$,  of the cases ($6/\pi^2$  is the probability
that two integers are relatively prime). We will remedy this below. }

Let us now  specialize to the particular example
of $X=T^4$. The partition function \single\ becomes
\eqn\teefouri{
\biggl(
\sum_{\Gamma^{4,4}}  q^{\half p_L^2} \bar q^{\half p_R^2}\biggr)
\Biggl(
{\vartheta_1(z\vert \tau)  \over  \eta}\Biggr)^2 {1\over  \eta^4}
{}~~
\overline{ \Biggl( { \vartheta_1(  z\vert \tau)  \over  \eta}\Biggr)^2 {1\over
\eta^4} }
}
Here $\Gamma^{4,4}$ is a lattice of zeromodes. 
We  will be interested in states with
zero $U(1)$ charges and as we discussed above this implies that
only states with $\tilde L_0 = 0 $ will contribute to the index we will
be computing. This implies that $p_R =0$ for each copy of the symmetric
product of $T^4$'s.  For generic values of the $T^4$ moduli this implies
that also  $p_L = 0$. If we go to the particular values where we have
additional values of $p_L$ allowed we see that they should appear in
pairs so that their contribution to the index cancels. We will
therefore drop the lattice sum in \teefouri .

Taking explicit derivatives and  using
the product formula:
\eqn\prodfrm{
\eqalign{
\vartheta_1(z\vert \tau)  &
= i (y^{1/2}-y^{-1/2})q^{1/8}
\prod_{n=1}^\infty (1-q^n)(1-y q^n)(1-y^{-1} q^n) \cr}
}
with $y = e^{2 \pi i z}$  gives:
\eqn\teefouri{
- \Biggl({\vartheta_1(z\vert \tau)  \over  \eta}\Biggr)^2 {1\over  \eta^4}
= \sum \hat c(n, \ell) q^n  y^\ell
}
The left hand side is a weak Jacobi form of weight
$-2$ and index $1$. Therefore, the coefficients
$\hat c(n,\ell)$ are actually functions of only one
variable $\hat c(n,\ell) = \hat c(4 n - \ell^2)$ \ez.
(This can also be seen by bosonizing the U(1) current.)
Using the sum formula
\eqn\sumform{
\vartheta_1(z\vert \tau) = i q^{1/8} (y^{1/2} - y^{-1/2})
\sum_{n=0}^\infty q^{n(n+1)/2} (-1)^n \chi_{j=n}(y^{1/2})
}
and $\chi_{j=n}(y^{1/2}) = \chi_{j=n/2}(y) + \chi_{j=(n-1)/2}(y)$ (valid for
$n>0$ and integral) and
expanding one easily derives explicit formulae
for the expansion coefficients:
\eqn\explicit{
\eqalign{
 \sum \hat c(n, \ell) q^n  y^\ell
& = (y^{1/2}- y^{-1/2})^2 \Biggl[ 1 
-2 (y^{1/2}- y^{-1/2})^2  q -  \cr
& (y^{1/2}- y^{-1/2})^2 (8 \chi_0 - \chi_{1/2} ) q^2 \cr
& - (y^{1/2}- y^{-1/2})^2 (24 \chi_0 - 8\chi_{1/2} ) q^3 + \cdots \Biggr] \cr}
}
Note that the positive powers of $q$ have an 
extra factor of $(y^{1/2}- y^{-1/2})^2$, and 
that $(y^{1/2}- y^{-1/2})^2= \chi_{1/2}- 2\chi_0$ 
allowing a decomposition into $SU(2)$ characters.

Finally, let us close with two remarks.
\item{1.} First, the expression $\CZ''$ in \second\ 
can be interpreted more geometrically. Recall that there is 
an action of $T^4$ on $\hilbk$ lifting the action 
by translation on $\symk$. The quotient space 
$\hiltil:= \hilbk/T^4$ is a simply 
connected irreducible hyperk\"ahler manifold. 
\foot{See, e.g. \beauville, and references 
therein. We thank N. Seiberg and 
E. Witten for discussions about 
this space.} Working in the charge zero sector the 
partition function factorizes:
\eqn\factorz{
Z(\hilbk) = Z(T^4) Z(\hiltil)
}
for $k\geq 1$. Therefore, 
\eqn\limihltl{
\eqalign{
\biggl( \sum_{k=1}^\infty p^k Z(\hiltil) \biggr)\vert_{\tilde y=1}
&= \lim_{\tilde y\rightarrow 1 } 
{\CZ(p,q,\bar q, y, \tilde y) - 1\over Z(T^4)}
=-\half {\eta^6 \over \vartheta_1^2} \CZ''  \cr}
}
so $\CZ''$ is essentially just the generating function 
for elliptic genera of the hyperk\"ahler spaces 
$\hiltil$. 

\item{2.} 
Second, similar multiplet counting formulae apply
to compactifications on $S^1 \times K3$. In this
case, $K3$ breaks half of the supersymmetries,
the BPS multiplets are smaller,
the  sigma model  is now $\sigma(\IR^4)
\times \sigma({\rm Sym}^k(K3))$ and the analog of
\deg\ is obtained by taking 
$\half {d^2 \over  d \tilde y^2}\vert_{\tilde y =
1}$ to get:
\eqn\kthrrdeg{
\eqalign{
(y^{1/2} - y^{-1/2})^2
\sum_{j_L, j_R}  \chi_{j_L}(-y)
 &  \chi_{j_R}(-1)
  D_{8/16}(Q_1,Q_5; j_L, j_R) \cr
+
 (y^{1/2} - y^{-1/2})^4 \sum_{N >0}
\sum_{j_L, j_R} q^N  \chi_{j_L}(-y) 
 &
\chi_{j_R}(-1)
  D_{4/16}^+(Q_1,Q_5,N; j_L, j_R) \cr
= (y^{1/2} - y^{-1/2})^2 {\prod (1-q^n)^4 \over
\prod_{n=1}^\infty (1- y q^n)^2 (1-y^{-1} q^n)^2}
&
{\Tr }_{S^k(K3)} [(-1)^F q^{L_0} {\bar q}^{\bar L_0} y^{2J_0^3}   ]
 \cr
= - (y^{1/2} - y^{-1/2})^4 {\eta^6(\tau)  \over
\vartheta_1(z \vert \tau)^2}
&
{\Tr }_{S^k(K3)} [(-1)^F q^{L_0} {\bar q}^{\bar L_0} y^{2J_0^3}   ]
 \cr}
}
We see that in this case the center of mass sigma model contributes
to the index for spacetime  BPS states.

\newsec{ $U$-duality and the Long String Interpretation}

 $U$-duality has interesting implications in
connection with
the long-string picture of \msu. The six dimensional 
$O(5,5;\IZ)$
$U$-duality group does not transform $N$, but, as
mentioned above can be used to put the string
charge $S$ in a canonical form, which we take to
be $S= (Q_1 Q_5, 1)$.
By a permutation like \ttprime\ we can then
map to  $S=(Q_1 Q_5, N)$. Then,  if $N$
and $Q_1 Q_5$ are relatively prime
we can again use
$U$-duality to  map to a charge vector of the
form $\gamma = (1,1;\vec 0; Q_1 Q_5 N)$.
This state is just  a single D1 and a single D5 with
momentum $N'= Q_1 Q_5 N$, and its degeneracy is the 
same as that of 
a  single long string. This implies that if we think
in terms of strings in the fivebrane \dvv , only the long string
contributes and all other contributions cancel.
It can be seen from \deg\ that indeed in this case only the
term with $s=1$ contributes to  \deg .

This description in terms of a long string applies when we take
$N$ to be coprime with $Q_1Q_5$. However, given $k=Q_1Q_5$ we should
consider all possible values of $N$ and the structure of the
Hilbert space is of the form:
\eqn\smhs{
\CH({\rm Sym}^k X) =\oplus_{ \{k_r\} }
\otimes_{r>0} {\rm Sym}^{k_r}(\CH_{r}(X)) 
}
where $\sum r  k_r =k$, and 
$\CH_{r}(X) $ is the single string Hilbert space
for a string of length $r$. The sectors which 
contribute to $\CZ''$ are of the form 
$ \oplus_{r \vert k} {\rm Sym}^{k/r} (\CH_{r}(X) ) \oplus
\cdots$ and  correspond to collections of strings of
a single length $k/r$. 

\subsec{A formula for all primitive vectors}

In this section we extend the counting formula from 
the case $gcd(Q_1,Q_5)=1$ to all primitive 
vectors equivalent to the three charge system.

$U$-duality
under the
transformation
$\CT$ \teeone\ is obvious. To check U-duality under  $\CT'$ we should 
remember that our formula is valid only if $gcd(Q_1,Q_5)=1$, therefore
we can compute  the right hand side of   \ttprime\
only if $gcd(N,Q_5) =1$ as well. In that case it is easy to see that
the sum over $s$ is such that the two results agree.
If one drops the restriction $gcd(Q_1,Q_5)=1$ then
\deg\ is {\it not} $U$-duality invariant.
As a simple example, let $p_1, p_2$ be two distinct primes. 
Then $\gamma = (p_1, p_1;0;p_2)$ on the RHS of 
\deg\ gives $\sum_\ell y^\ell \hat c(p_1^2 p_2, \ell)$
while $\gamma = (p_2,p_1;0;p_1) $ gives 
$\sum_\ell y^\ell \hat c(p_1^2 p_2, \ell) 
+ \sum_\ell p_1 y^{\ell p_1}  \hat c(p_1 p_2, \ell)$.

To cure this problem we begin by noting that, 
in close analogy to the
remark of \dmvv,  the expression on the RHS of \deg\ is
just a transform by a Hecke operator 
 $V_{Q_1 Q_5}$ applied 
to  a Jacobi form \ez.
\foot{ The power of $s$ in \deg\ is nonstandard. 
A weight $=k$ form would have a power of $s^{k-1}$. 
In our case, $\hat c(n,\ell)$ are coefficients 
of a weight $k=-2$ form, but the power of $s$ in 
\deg\ corresponds to a weight $k=+2$ form. 
We do not understand this peculiarity very well, 
but it does not affect the following argument. } 
Since $V_{Q_1 Q_5} = V_{Q_1} V_{Q_5}$  for
$Q_1, Q_5$ relatively prime one might wonder if the
general formula is given by $V_{Q_1} V_{Q_5}$.
Indeed, this guess has some very attractive 
features. Following \ez, pp. 44-45 one can write
(for any $Q_1, Q_5$, not necessarily relatively 
prime):
\eqn\hecke{
(V_{Q_1} V_{Q_5}\hat c)(N, \ell) = 
\sum_s N(s)~  s~  \hat c({N Q_1 Q_5 \over s^2}, {\ell \over s})}
where $N(s)$ is the number of integral divisors 
$\delta$ of 
\eqn\heckeii{
N, Q_1, Q_5, s, {N Q_1 \over s},{N Q_5 \over s},
{ Q_1 Q_5 \over s},{N Q_1 Q_5 \over s^2}
} 
It follows that  $V_{Q_1}V_{Q_5} = V_{Q_5} V_{Q_1}$ 
and, more importantly, that  
 \hecke\ is {\it completely symmetric} in 
$Q_1, Q_5, N$. Thus, 
this is a natural $U$-duality 
invariant ansatz for the general case. 
Indeed, it is the unique $U$-duality invariant 
extension to primitive 3-charge systems. 

\newsec{Comparison to Supergravity}

In this section we compute $\CE_2$ by summing over multiparticle
supergravity excitations of $AdS_3\times S^3\times T^4$ and 
using the AdS/CFT correspondence \jmone. A similar
comparison with the elliptic genus for the $K3$ case was made by
de Boer \dbr.

The supergravity computation is in the  NS sector while the CFT
partition function is normally calculated in the R sector. Under
spectral flow between the sectors a state with weight $h_R$ and
(half-integer) 
U(1) charge $j_R$ is mapped into a state with weights \eqn\map{
h_{NS} = {k\over 4 } + h_R + {j_R  }~~~~~~~~~~ j_{NS} = j_R +
{k\over 2 } } where $k = c/6$ ($c=6Q_1Q_5$ is the central charge
of the CFT). We use a convention such that $h_{NS} =-k/4 $ for the
NS vacuum. The partition function in the NS sector can be obtained
from the partition function in the R sector by the following
replacements \eqn\repl{ p \to pq^{1/4} y ~,~~~~~ q \to q ~,~~~~~~~
y \to  y q^{1/2} }

In principle one expects agreement with supergravity only for
small conformal weights, not much bigger than the NS vacuum
$h_{NS} = - k/4 $. When conformal weights are  of order $k$ the stringy
exclusion principle \msad\ is relevant and supergravity breaks
down.   We shall in fact find agreement for all negative values of
$h_{NS}$, $-k/4 \leq h_{NS} <0$.

For the CFT we start from \final\ in the RR sector and
we find the NS-NS partition function
 \eqn\fsqns{
{1 \over 2}\CZ_{NS}'' = \sum_{n,m,l,s} 
\hat  c(4nm-l^2) s \left( p^n q^{n/4 +m
+l/2} y^{n+l} \right)^s .
}

Now we concentrate on the terms in this expansion with negative
powers of $q$, relevant for the comparison to supergravity. The
only possibility is $n=1,~m=0,~l=-1$, since $\hat c(r) =0$ for
$r< -1$. Using $\hat c(-1) = 1, ~
\hat c(0) =-2 $ this gives 
\eqn\finalres{ {1 \over 2 } \CZ_{NS}'' = \sum_s    s (
p q^{-1/4} )^s + . ~.~.~ }
where the dots involve non-negative powers of $q$.

Now we consider the supergravity calculation. 
We need to define an appropriate notion of a 
``supergravity elliptic genus'' $Z_{\rm sugra}(p,q,y)$. 
We will follow the proposal of de Boer \dbr. 
The single particle supergravity  Hilbert space can
be derived by group theory and Kaluza-Klein reduction.
It decomposes as a representation of
$SU(2\vert 1,1) \times SU(2 \vert 1,1)$:
\eqn\decomps{
\CH_{\rm single\ particle}
= \oplus_{j, \tilde j\geq 0} N_{j, \tilde j} ~~ (j) \otimes
(\tilde j)
}
Short $SU(2\vert 1,1)$ reps
 are labelled by the maximal spin, i.e.,
a nonnegative half-integer $j$. The highest weight
has  $h= j$. Label it by $(j)$.
It turns out that single particle states are always
products of short representations. There is
no analog of the $ long \otimes short $ of CFT. 
(These latter come from multiparticle 
supergravity states.)
It turns out that the degeneracies in
 \decomps\ can be read  off from
the identity of \gottschsoergel\ 
\eqn\gottsche{
\sum_{k\geq 0} p^k \sum_{r,\tilde r} h^{r,\tilde r}(S^k X) y^{r}
 \tilde y^{\tilde r}
=
\prod_{n=1}^\infty \prod_{r,\tilde r} (1-(-1)^{r+\tilde r} y^{r+n-1} \tilde
y^{\tilde r+n-1} p^n)^{-(-1)^{r+\tilde r} h^{r,\tilde r}(X)}
}
where $h(X)= \sum_{r, \tilde r}(-1)^{r+\tilde r}
 h^{r,\tilde r} y^r \tilde y^{\tilde r}$ 
is the Hodge polynomial. 
The generating function \gottsche\ counts
 $(c,c)$ primaries. Each $(c,c)$ primary 
 in turn corresponds to  a
short $SU(2\vert 1,1) \times SU(2 \vert 1,1)$
representation. 
De Boer \dbr\   proposes to associate a new quantum
number to the supergravity states, the degree,
in order to take into account the exclusion principle.
The degree 
 is the power of $p$ multiplying the various
factors in \gottsche.
Thus, representations  are now labelled by
$(r,\tilde r;d)$ where $d$ is the degree.
Notice that this assignment of degree breaks the 
$SO(4,5)$ continuous U-duality symmetry of supergravity 
on $AdS_3\times S^3 \times T^4$. 

With this innovation the single-particle Hilbert space is:
\eqn\singlpart{
\CH_{\rm single\ particle} = \oplus_{n\geq 0, r,\tilde r} h^{r,\tilde r}(X)
(\half(n+r),\half(n+\tilde r);n+1)
}
where $h^{r,\tilde r}(X) $ are the Hodge numbers of
$X=K3, T4$.
For the torus the Hodge  polynomial factorizes
as $(1-y)^2(1-\tilde y)^2$ so we can introduce the
useful device for the torus Hilbert space:
\eqn\singlpartee{
\CH_{\rm single\ particle} = \oplus_{n\geq 0}
\oplus_{r,\tilde r=0,1,2}
 d(r) d(\tilde r)  (\half(n+r),\half(n+\tilde r);n+1)
}
Here $d(0)=d(2) = 1, d(1)=-2$ (the sign is for a 
fermionic representation). Notice that we are including 
the identity. 

We now  define the
``supergravity elliptic  genus'' as the free
field theory partition function for the
Fock space built up from
$\CH_{\rm single\ particle} $:
\eqn\sugraeg{
\CZ_{\rm sugra}(p, q, \tilde q, y, \tilde y)  :=
 \prod_{\CH_{\rm single\
particle}}
(1-p^d q^\Delta \tilde q^{\tilde \Delta} y^{\ell} \tilde y^{\tilde \ell}
)^{-(-1)^{\ell + \tilde \ell}}
}
(here it is more convenient to use $\ell = 2j$ which is 
integral). 
Since we will eventually set $\tilde y=1$ and 
expect only holomorphic quantities from 
left chiral primaries we will 
temporarily suppress $\bar q$. This is not 
totally innocent, and we  will return to the
$\bar q$-dependence at the end of this section. 
Suppressing $\bar q, \tilde y$, 
we can rewrite \sugraeg\ as a product 
of factors
$(1-p^n q^h y^\ell)^{ - c_s(n,h,\ell) } $ where  $c_s(n,h,\ell)$ is the
number of single particle states with $L_0 = h$, 
$U(1)$ charge  $=\ell$ and 'degree' $n$. (As usual $c <0$ for fermions).
Here we are
measuring $L_0$ relative to the NS vacuum, as is conventional in
$AdS$ discussions. So in order to compare with the above formulae
we need to replace $p \to p q^{-1/4}$.  
The effects of the exclusion principle are  approximated by 
truncating the supergravity spectrum to states with total 
degree $k=Q_1Q_5$.

The full exclusion-principle-modified supergravity partition function
is thus 
\eqn\fullsugra{\CZ_{sugra} =
 \prod_{n,h,\ell} { 1 \over(1-p^n q^{h-n/4} y^\ell)^{  c_s(n,h,\ell) } }
} 
Of course, as written $\CZ_{\rm sugra}=1$ 
for $T^4$ at $\tilde y=1$. 
We therefore  need to put back
 $\tilde y$ and take derivatives to get a
nontrivial quantity.  The $\tilde y$ that appears in the R partition
function differs from the one appearing in the NS partition
function by a factor of  $ {\bar q}^{1/2}$ arising in the spectral flow.
So after differentiating twice we set $\tilde y = {\bar q}^{1/2}$. This
selects the chiral primaries. Manipulations similar to those in
section five then lead to
 \eqn\sugrare{ \CZ_{sugra}''|_{\tilde y
= {\tilde q}^{1/4} } = \sum_{s,h,n,\ell} \hat c_s(n,h,\ell) s ( p^n
q^{h-n/4} y^\ell)^s } where $\hat c_s(n,h,\ell) = \sum_{\tilde \ell}
{\tilde \ell}^2
 c_s(n,h,\ell,\tilde \ell)$ counts the number of right chiral primaries with the
 given properties
and the sum over $\tilde \ell$ runs over all the chiral primaries of
degree $n$.

Next we need a good way to enumerate chiral primaries in this
theory. Using \singlpartee\  above we 
 can perform the sum over ${\tilde \ell }^2$ over chiral
primaries of given degree. It is easy to see that $\sum_r d(\tilde
r) = 0$, $\sum d(\tilde r) ( n + \tilde r) =0$ and $\sum_r
d(\tilde r) (n+\tilde r)^2 = 2$. This final sum is independent of
$n$ and just gives an overall factor, as in the CFT result. We now
need to compute $c(n,h,\ell)$ just for the left-moving piece.
Ignoring for a moment the sum over $s$ we see that we have
\eqn\sumsugra{ \sum c(n,h,\ell) p^n q^{h-n/4} y^\ell =
 \sum_{n,k,r,t} d(r) d(t)p ^{n+1}
q^{ { n+ 2r +2t -1\over 4} + k} \sum_{\ell=-{n +r - t}}^{n+r-t} y^\ell }
where $r=0,1,2$ as above and $t=0,1,2$ takes into account the
descendants of the form $G_{-1/2}$, etc. The sum over $k$ takes
into account the descendants of the form $L_{-1}^k$, $k =0,1,2$. The
sum over $\ell$ is in steps of 2. We have replaced $p \to p
q^{-1/4}$ to take into account the ground state energy so that we
can compare to \finalres . The sum over $s$ is taken into account
by replacing $(p,q,y) \to ( p^s, q^s,y^s) $ multiplying by $s$ and
then summing over $s$. We are interested in terms with negative
 powers of $q$. This requires
  \eqn\cond{
  { n-1 \over 4 } + {r + t \over 2 } + k < 0
} The only possibility is  $k=r=t=n=0$, and this reproduces
\finalres. Hence the supergravity and CFT calculations of $\z$
agree exactly for all negative powers of $q$.
Notice that basically only the ground state is contributing to
\finalres . So the agreement boils down to the statement that
all the gravity contributions cancel at low enough energies. 

It is not hard to see that the agreement does not 
persist for nonnegative powers of $q$ 
(indeed, there is a discrepancy at order 
$q^0$).  This is not surprising because 
supergravity becomes strongly coupled before this point. 
Indeed a black hole which is a left-chiral primary appears at this level.
This black hole is an extremal rotating black hole with angular momentum
on $S^3$. 

Finally, let us return to the issue of the $\bar q$ dependence of 
$\CZ_{\rm sugra}''$. 
In fact if $\bar q$ 
is reinstated, one finds at these excited levels dependence on 
positive powers of $\bar q$. This might seem to be 
a contradiction because we argued in section 3 that the large 
$\CN=4$ algebra forbids  $\bar q$-dependence of $\z$. 
What happens is that the implementation of the exclusion principle 
as a cutoff on supergravity states breaks  the large 
$\CN=4$, which for example maps single particle states below the cutoff to 
multi-particle states above the cutoff. Hence this implementation,
while very successful at low energies,  
is too naive to describe the Hilbert space at high energies. 
Indeed, the $\bar q$ dependence at order $p^N$ first shows up at 
order
$\bar q^{N/4}$. Thus, in the large $N$ limit the action of the 
large $\CN=4$ algebra is restored, in accord with the 
AdS/CFT correspondence. 

\newsec{Open Questions}

As we have stressed,  \deg\  is false when $Q_1, Q_5$ have common
factors, i.e., when the Mukai vector of the instanton moduli space
is not primitive. This is not terribly surprising since it is
known that the moduli space is singular under such circumstances,
and there are even resolutions of the space not equivalent to the
Hilbert scheme of points \ogrady. Physically, nonprimitive vectors
are associated with the possibility of boundstates at threshold so
we expect subtleties in counting BPS states. Very similar
subtleties  were found already in the work of Vafa and Witten in
\vw. In view of this one should be cautious about the existing
formulae for BPS states in $S^1 \times K3$ compactification for
nonprimitive Mukai vectors. Unfortunately, $U$-duality is not a
useful tool for probing this question.

In \hecke\heckeii\ we extended \deg\ to all 
primitive 3-charge systems, but this leaves open 
the question of what 
the degeneracies really are when $\gamma$ is not 
primitive.   Because
of bound-states at threshold this question requires careful
definition. One way to approach this question 
is to use the trick in \sen, compactifying on another circle
and turning on a charge to remove the boundstates at threshold.

The results of this paper
raise some interesting open problems. It
is  natural to expect that the full set of
BPS states for toroidally compactified
type II string  is counted by some interesting automorphic
functions transforming nontrivially
 under $E_{d,d}(\IZ)$.
One might hope that such forms  might appear in quantum
corrections along the lines of the BPS counting formulae appearing
in quantum corrections. (See \po, p.10 for a list of 
references.) 
At present such automorphic forms remain part of
the Great Unkown.

Finally it would be interesting to compute
this index for supergravity on $AdS_3 \times S^3 \times S^3
\times S^1$ \adssss\ in order to see what we can learn about the
conformal field theory.

\bigskip
\centerline{\bf Acknowledgements}\nobreak
\bigskip

We would like to thank R. Dijkgraaf, B. Gross, 
N. Seiberg,  C. Vafa and  E. Witten
for very
helpful discussions and correspondence.
GM is supported by DOE grant
DE-FG02-92ER40704 and  the Monell 
Foundation. JM and AS are supported by DOE grant
DE-FG02-92ER40559. JM is also supported by 
NSF PHY-9513835, the Sloan Foundation and the David and Lucille 
Packard Foundation. JM is also a   Raymond and 
Beverly Sackler Fellow.

\listrefs

\bye